\documentclass{raa}
\usepackage{graphicx,times}

\begin{document}

\title{Gap formation in a self-gravitating disk and the associated
migration of the embedded giant planet}

\author{Hui Zhang\inst{1}, Huigen Liu\inst{1}, Ji-Lin Zhou\inst{1},Robert A. Wittenmyer\inst{2}}
\institute{School of Astronomy and Space Science \& Key Laboratory of
Modern Astronomy and Astrophysics in Ministry of Education, Nanjing
University, Nanjing 210093,China; {\it huizhang@nju.edu.cn}\\
\and
Department of Astrophysics, School of Physics, University of NSW, 2052, Australia\\
}

\abstract{We present the results of our recent study on the
interactions between a giant planet and a self-gravitating gas disk. We
investigate how the disk's self-gravity affects the gap formation
process and the migration of the giant planet.
Two series of 1-D and 2-D hydrodynamic simulations are performed. We
select several surface densities and focus on the gravitationally stable
region. To obtain more reliable gravity torques exerted on the planet, a
refined treatment of disk's gravity is adopted in the vicinity of the
planet.
Our results indicate that the net effect of the disk's self-gravity on the
gap formation process depends on the surface density of the disk. We
notice that there are two critical values, $\Sigma_I$ and $\Sigma_{II}$.
When the surface density of the disk is lower than the first one, $\Sigma_0
< \Sigma_I$, the effect of self-gravity suppresses the formation of a gap.
When $\Sigma_0 > \Sigma_I$, the self-gravity of the gas tends to benefit the
gap formation process and enlarge the width/depth of the gap. According
to our 1-D and 2-D simulations, we estimate the first critical surface
density $\Sigma_I \approx 0.8MMSN$. This effect increases until the
surface density reaches the second critical value $\Sigma_{II}$. When
$\Sigma_0 > \Sigma_{II}$, the gravitational turbulence in the disk
becomes dominant and the gap formation process is suppressed again. Our
2-D simulations show that this critical surface density is around
$3.5MMSN$. We also study the associated orbital evolution of a giant
planet. Under the effect of the disk's self-gravity, the migration rate
of the giant planet increases when the disk is dominated by
gravitational turbulence. We show that the migration timescale
associates with the effective viscosity and can be up to $10^4 yr$.
}

\authorrunning{H. Zhang et al. }            
\titlerunning{Gap formation in a self-gravitating protoplanetary disc}  
\maketitle

\section{Introduction}

To date, more than 900 exoplanets have been confirmed. The great
diversity in the orbital characteristics of exoplanets reveals
complicated physical and dynamical processes in the formation and
evolution of exoplanets. One of the most important dynamical processes
is the interaction between exoplanets and the protostellar disk in which
they are embedded. The physical properties of the protostellar disk
usually dominate the initial conditions of the subsequent orbital
evolution of the exoplanet system. Thanks to the improvement of direct
imaging methods, a number of protostellar or debris disks interacting
with exoplanets have been resolved, e.g. Fomalhaut(\cite{kal08}) and HR
8799(\cite{mar08}). Their detailed structures, such as the gaps created
by the embedded planets, may be revealed in the near future.

According to the general theory of disk-planet interaction, a planet
embedded in a protostellar disk will generate density waves in the disk.
For a planet of a few Earth masses ($M_\oplus$), the response of the
disk is linear and the structure of the disk is almost
unchanged(\cite{gol79,war97}). On the other hand, for a planet with mass
comparable to that of Jupiter, the response of the disk becomes
nonlinear and it usually results in a density gap at the position of the
planet's orbit. In this regime, the planet is locked and moves as a part of the disk,
which is called the type II migration (\cite{lin86}).

The gap formation process is a key issue to understand the type II
migration. In an inviscid disk, the gravitational tidal force exerted on
the gas by a giant planet tends to split the disk, while the local fluid
pressure resists the creation of any low density region. So, the
criterion for gap formation is that the planet's Roche radius exceeds
the pressure scale height of the disk. In the case of a viscous disk,
the dissipation driven by the viscosity of the gas also tends to
replenish the gap. As a result, the gap formation condition usually
depends on the planet-star mass ratio $M_p/M_\ast$, the semi-major axis
of the planet's orbit $a_p$ , the scale height of the disk $H$ and the
viscosity of the gas $\alpha$(\cite{lin93}):
\begin{equation}
\frac{M_p}{M_\ast} \equiv 40\alpha (\frac{H}{a_p})
\end{equation}
However, to determine the width of a gap is not straightforward.
Basically, the width of a gap is determined by the wave propagation
length scale, and should be a decreasing function of the effective viscosity
of the gas (\cite{lin93};\cite{tak96}). One may be rigorously define the positions of gap boundaries as the places where the tidal torque of the planet balances the torque
raised by the viscous stress, given that all the other effects have
already achieved equilibrium, e.g. the gravity of the central star, the
thermal pressure and the centrifugal force of the gas. While most of
these factors turn out to be strongly coupled with the surface density
profile. If the perturbing planet is small, the response of the disk is linearly analysable. However, a Jupiter size planet as we consider here cannot
be treated as a small perturbation. The density waves it excites are
shocks and the associated gap formation process is a highly nonlinear
process. Thus, numerical simulation is still the most powerful tool to
study this process.

Many simulations have been performed to investigate the gap formation
process in laminar viscous disks(\cite{tak96};\cite{kle99};\cite{lub99};
\cite{ang03}) and in MHD turbulent disks(\cite{win03};\cite{pap04}). There is an important issue that has been investigated poorly so far, which is the
self-gravitating effect of the gas (disk). In a non-ionized disk,
gravitational turbulence is the most important source of the effective
viscosity. In a high density self-gravitating disk, the gravitational
turbulence can be very strong and even a Jupiter size planet may not be
able to open a gap(\cite{bar11}). Conversely, in a low density disk, the
self-gravitating effect is usually neglected or just treated as an
effective viscosity(\cite{gam01}). So, it seems that as the surface
density of the disk increases, the self-gravitating effect will only
result in higher effective viscosity and monotonically reduce the gap size. When
the density is high enough, even a giant planet could not open a gap.

However, the self-gravity potential is in fact coupled with the
equilibrium angular velocity of the gas. As the self-gravity potential
varies with the surface density profile, the angular velocity required
by the equilibrium varies as well. Thus, the gas needs to drift inward
or outward to achieve a new equilibrium, especially at the boundaries of
the gap where the self-gravity potential changes the most. As a result,
the self-gravity of the gas changes the size of the gap and its net
effect on the gap formation process may not be straightforward, and
systematic numerical experiments are needed. In this paper, we focus on
the gravitationally stable region of the disk's surface density and
investigate how the self-gravitating effect really affects the gap
formation process, as well as the subsequent migration of the embedded
giant planet.

We perform both 1-D and 2-D simulations to investigate the disk-planet
interactions with the disk's self-gravity effect included. Our results
show that the self-gravity does not suppress the gap formation process
monotonically. Instead, there are two critical surface densities
$\Sigma_I$ and $\Sigma_{II}$. When $\Sigma_0 < \Sigma_I$, where
$\Sigma_0$ is the initial surface density of disk, the gap formation
process is suppressed when the self-gravitating effect is included. When
$\Sigma_0 > \Sigma_I$, the self-gravitating effect benefits the gap
formation process and results in a wider gap. This enlargement enhances
until the second critical surface density $\Sigma_{II}$
is reached. When $\Sigma_0>\Sigma_{II}$, the gravitational turbulence
viscosity becomes dominant in the disk and the gap formation process is
suppressed again. The exact value of this two critical densities may depend on the many physical settings. In our simulations, we use $MMSN$(the surface density in the minimum mass Solar nebula model\cite{hay81}) as a unit of disk surface density. Then, the first critical density is around $0.8$ and the second one is around $3.5$. The associated migration of the giant planet is also studied and we find that the self-gravity of gas accelerates the type II migration when $\Sigma_0>\Sigma_{II}$. We confirmed that the migration time scale associates with the effective viscosity in the disk, and can be as
short as $\sim 10^4 yr$ in a very dense disk $\Sigma_0\geq7 MMSN$.

This paper is arranged as follows: We introduce the models of the 1-D
and 2-D simulations in Section 2. The results are described in section
3. In section 4 we summarize our conclusions and discussions. The
details of refined treatment of the gravity torques and the calculation
of the disk's self-gravity are described in Appendix \ref{torque} and
\ref{sg}.

\section{Numerical Model}

\subsection{Computational Units}

To normalize our calculations, we set the mass of the central star to be
the mass unit $M_\ast=1$ and the gravitational constant $G=1$. The
length unit is set to be the initial orbital radius of the planet
$a_0=1$. Thus the orbit frequency of the planet is unity and its initial
orbital period is $P_0=2\pi$. According to this
configuration, our scale is in fact arbitrary. To connect with the real
physical dimensions, we further set the central star to be one Solar
mass $M_\ast=M_\odot$ and the initial orbital radius of the planet is
$a_0=5.2AU$. Thus the time unit becomes $11.2yr/2\pi$. According to the
minimum-mass Solar nebular model(MMSN,\cite{hay81}),
\begin{equation}
\Sigma = \Sigma_0 (\frac{a}{1AU})^{-3/2},
\end{equation}
where $\Sigma_0=1700 g cm^{-2}$. According to our length unit, where
$a_0=5.2AU$, the density constant in our model is $\Sigma_0\approx140 g
cm^{-2}$. To be convenient, we set $MMSN=140 g cm^{-2}$ as a surface
density unit in this paper. Therefore, $2MMSN$ equals to
$\Sigma_0\approx280g cm^{-2}$ at $a_0=5.2AU$.

\subsection{1-D model}

In the first series of simulations, we solve the viscous evolution of a
1-D self-gravitating disk which is perturbed by a Jupiter mass planet.
Assuming $\Sigma$ is the surface density of the disk, $\Omega$ and $v_r$
is the angular and radial velocity of gas, the 1-D continuity equation
is
\begin{equation}
r\frac{\partial\Sigma}{\partial t}+\frac{\partial}{\partial r}(r\Sigma v_r)=0,
\end{equation}
and the equation of angular momentum reads:
\begin{equation}
r\frac{\partial(\Sigma r^2\Omega)}{\partial t}+\frac{\partial}{\partial r}(r\Sigma v_r\cdot r^2\Omega)=\frac{1}{2\pi}\frac{\partial \tilde{G}}{\partial r},
\end{equation}
where $\tilde{G}$ is the transportation rate of angular momentum.
By eliminating $v_r$, we obtain the governing equation:
\begin{equation}
\frac{\partial \Sigma }{\partial t}=-\frac{1}{r}\frac{\partial}{\partial
r}[(\frac{1}{2\pi}\frac{\partial \tilde{G}}{\partial r}-r^2\Sigma \frac{\partial v_\theta}{\partial t})/(\frac{\partial
rv_\theta}{\partial r})]
\end{equation}
where $v_\theta=r\Omega$.

The transportation rate of angular momentum $\tilde{G}$ is mainly determined by two factors: the effective viscosity of the gas
$\tilde{G_{\rm \nu}}$ and the torques exerted by the planet $\tilde{G_{\rm p}}$. For the first factor,
\begin{equation}
\tilde{G_{\rm \nu}}=2\pi r\cdot \nu \Sigma r\frac{d\Omega}{dr} \cdot r,
\end{equation}
where $\nu=\nu_{art}+\nu_{sg}=(\alpha_{art}+\alpha_{sg}) (H/r)^2_s \Omega$. $H/r$ is the scale height ratio of the disk. To identify the contribution from the disk's self-gravity, the total viscosity is divide into two parts: $\alpha_{sg}$, the effective viscosity caused by the self-gravitating effects and $\alpha_{art}$, an artificial viscosity denoting viscosity comes from all other effects, e.g. the MRI. The typical value of $\alpha_{art}$ ranges from $10^{-3}$ to $10^{-2}$. We are considering a self-gravitating disk, most of viscosity is assumed to be caused by the self-gravitating effect, so we adopt a low artificial viscosity: $\alpha_{art}=10^{-3}$.

When a planet is embedded in the disk, its tidal torques causes additional
angular momentum transportation(\cite{lin79a}).
The total transportation rate of angular momentum becomes:
\begin{equation}
 \tilde{G}=2\pi r\cdot \nu \Sigma r\frac{d\Omega}{dr}+\tilde{G_{\rm p}},
\end{equation}
where $\tilde{G_{\rm p}}$ is the torque exerted on the gas by the embedded planet, which contains both the linear
Lindblad and corotation torques for isothermal gas(\cite{paa10}). Following the Eq(14) and Eq(15) of \cite{war97}, we can obtain the
smoothed lindblad torque density. In corotational region, the linear corotational torque density is represented in Eq(16) of
\cite{paa10}. Hence, we can finally obtain the torques $\Gamma_{\rm r}$ as well as the torque density $\frac{\partial \Gamma_{\rm
r}}{\partial r}$ on the gas at radius $r$ duo to the planet.


The self-gravitating effects are simulated by two terms: the
self-gravitating viscosity $\alpha_{sg}$ and the self-gravity potential
on the disk $\Phi_{sg}$. Since the 1-D model could not simulate the
gravitational turbulence well, we adopt an analytic description of
$\alpha_{sg}$. We will discuss it in detail in Section 3.2.1. Besides
the effective viscosity, the self-gravity of the disk $\Phi_{sg}$ may
also change the meridional velocity field $v_{\theta}$ on the disk.
Considering this as a quasi-static process, we have:
\begin{equation}
\frac{v_\theta^2}{r}=\frac{GM_\ast}{r^2}+\frac{1}{\Sigma }\frac{dP}{dr}+\frac{d\Phi_{sg}}{dr}.
\end{equation}
In this 1-D model, $\Phi_{sg}$ is calculated by integrating the radial
component of the gravity of all the grids on the disk. To avoid singularity, a softening length $\epsilon=0.1H$ is adopted. We emphasize that the self-gravity potential
$\Phi_{sg}$ is not constant, instead, it changes every time step as the
surface density changes. So as the equilibrium angular velocity. This
effect may imply a size change of the gap.

The governing equation is diagonalized to a tridiagonal matrice. Methods to solve this kind of linear algebraic equations can be found in\cite{nr92}. The initial surface density $\Sigma_0$ equals to a series of values: from $0.7MMSN$ to $2.8MMSN$. The boundary condition is set to be solid where $\Sigma_{bound}=\Sigma_0$.

\subsection{2-D model}

In the second series of simulations, we solve the vertical integrated
continuity and momentum equations in a 2-D cylindrical coordinates by
our ANTARES code. The details and convergence tests of our code can be
found in Zhang \textit{et al.} (2008) and Zhang \& Zhou (2010a),
respectively.

\subsubsection{Numerical Method}

We assume the disk is thin and cold, where $H/r=0.02$. The vertically
averaged equations are solved in a 2-D cylindrical coordinates
$(r,\theta)$, whose origin is located at the central star. To make sure
each cell is almost square, we adopt a logarithmic grid along the radial
direction with a constant ratio $\beta=\Delta r/(r\Delta\theta)\approx0.8$.

The major difficulty of numerical experiments on the self-gravity effect
is poor computational efficiency. It is too time-consuming to solve the
Poisson equation of the gravity potential on a highly perturbed disk,
where the density varies quickly with both time and position. Thanks to
the application of the Fast Fourier Transform method, we could greatly
reduce the complexity of this problem from $N^2$ to $N\ln N$, where $N$
is the total number of the grids used to resolve the disk. Despite this
improvement, it is still too ``expensive'' to perform high resolution
2-D or 3-D simulations, when the total $N=N_r\times N_\theta > 10^6$. On
the other hand, low resolution simulations usually introduce un-physical
effects and the results are thus less reliable.

One of the most significant numerical effects on an Eulerian grid is
that, since the mass is placed at the center of each cell instead of
smoothly spreading over it, the net gravity force exerted on the planet
is usually dominated by the mass within only the single cell whose
center is immediately adjacent to the planet. As the planet travels
through a series of cells, the net torque exerted on it experiences
un-physical large variations. When the resolution of the grid is high
enough, this effect could be partly reduced by a well-chosen softening
length. However, choosing the value of the softening length is
difficult. On one hand, it should be small. It is usually smaller than
the scale height of the disk or the Hill radius of the planet. On the
other hand, it needs to be large enough that the softening region can be
resolved by the grid size. It usually requires large number of grids to resolve the immediate vicinity of the planet, e.g. the corotation zone of the planet(\cite{mas04}). This necessitates high grid resolution as well. To balance the computational efficiency
and accuracy, we adopt a relatively low mesh resolution $N_r\times
N_\theta=256\times512$ and a refined treatment of the gravity torque in
the vicinity of planet(see Appendix \ref{torque}).

The velocity is denoted by $v=(v_r,v_\theta)$, where $v_r$ is the radial
velocity and $v_\theta$ is the velocity in the azimuthal direction. The
vertically averaged continuity equation is given by
\begin{equation}
\frac{ \partial \sigma }{ \partial t} + \frac{1}{r}\frac{
\partial(r \sigma v_r) }{ \partial r } + \frac{1}{r} \frac{ \partial(
\sigma v_\theta)}{ \partial \theta }=0
\end{equation}
The momentum equations in the radial and azimuthal directions are
\begin{equation}
\frac{ \partial ( \sigma v_r ) }{ \partial t}+\frac{1}{r} \frac{
\partial(r \sigma v_r^2) }{ \partial
r}+\frac{1}{r} \frac{ \partial( \sigma v_r v_\theta)}{ \partial
\theta }=\sigma \frac{v_\theta^2}{r}-\sigma \frac{ \partial \Phi
}{
\partial r}-\frac{ \partial P }{ \partial r }
\end{equation}
\begin{equation}
\frac{\partial ( \sigma v_\theta )}{ \partial t
}+\frac{1}{r}\frac{
\partial(r \sigma v_r v_\theta)}{ \partial r}+\frac{1}{r} \frac{
\partial( \sigma v_\theta^2)}{ \partial \theta}=-\sigma \frac{v_r v_\theta}{r}-\sigma
\frac{ \partial \Phi}{\partial \theta}-\frac{1}{r} \frac{ \partial
P}{ \partial \theta }
\end{equation}
The external potential $\Phi$ is:
\begin{equation}
\Phi = \Phi_S +\Phi_p + \Phi_D + \Phi_{N,p} + \Phi_{N,D}
\end{equation}
where $\Phi_S = -GM_{\odot}/|\vec{r}|$ is the potential of the central
star, $\Phi_p = -G M_p/(|\vec{r}-\vec{r}_p|+\varepsilon) $ is the
potential of the planet, $\Phi_D$ is the self-gravity potential of the
gaseous disk, which is determined by the Poisson equation:
\begin{equation}
\nabla^2\Phi_D = 4\pi G\Sigma,
\end{equation}
In our simulation, we calculate the self-gravity force $F_{sg}$ directly
by the FFT method (see Appendix \ref{sg}). $\Phi_{N,p}$ is the indirect
potential caused by the Jupiter-mass planet:
\begin{equation}
\Phi_{N,p} = \frac{M_p}{M_\odot + M_p} {\Omega_p}^2 \vec{r}\cdot\vec{r}_p
\approx \frac{G M_p}{{r_p}^3} \vec{r}\cdot\vec{r}_p,
\end{equation}
$\Phi_{N,D}$ is the indirect potential due to the gravity of the gas disk:
\begin{equation}
\Phi_{N,D}=G\int_D\frac{\vec{r}\cdot\vec{r'}}{|\vec{r'}|^3}dm(\vec{r'}).
\end{equation}
Since we have assumed the disk is very cold and we focus on the gravitational stable region, we do not adopt the energy equation in the 2-D model. Instead, we adopt a locally isothermal equation of state:
\begin{equation}
p=\Sigma c_s^2
\end{equation}
where $c_s$ is the sound speed which is only the function of $r$: $c_s=(H/r)v_{kep}$ and $v_{kep}=\sqrt{GM_\ast/r}$ is the local Keplerian velocity. We do not employ any artificial viscosity, however the numerical viscosity due to the coarse grid is $\nu_{num}\sim10^{-5}$.

To estimate the numerical viscosity we have performed several short-time simulations to test the diffusion time of a density ring in the disk under different resolutions: $256\times512$, $512\times1024$, $1024\times2048$,$1600\times3200$ and $2048\times4096$. The planet and disk's self-gravity are not included. We found that the diffusion time doesn't change anymore when we change the resolution from $1600\times3200$ to $2048\times4096$. We believe that the grid effect is neglectable when the resolution achieves $2048\times4096$. Then we add an artificial viscosity in the $2048\times4096$ case. When this artificial viscosity increases to $10^-5$, we found the diffusion time is comparable to the value of the $256\times512$ case. So, we conclude the viscosity comes from the coarse grid is about $10^-5$ in a resolution of $256\times512$."

\subsubsection{Initial and Boundary Conditions}

We fix the star at the origin of the frame and let gas and the planet
travel around it. The initial orbit of the planet is circular and its
semi-major axis is set to be unity, $a_0=1$. To ensure the gas disk
starts with an equilibrium state, the initial azimuthal velocity field
is set to be $v(r)_{\theta 0}=(1/r+r F_{sg}(r)-c_s(r)^2)^{1/2}$, where $F_{sg}(r)$ is the self-gravity of the gaseous disk and $c_s(r)$ is the local sound speed. The initial radial velocity of gas $v_{r0}$ is set to be $0$.

\begin{figure}
  \includegraphics[width=140mm]{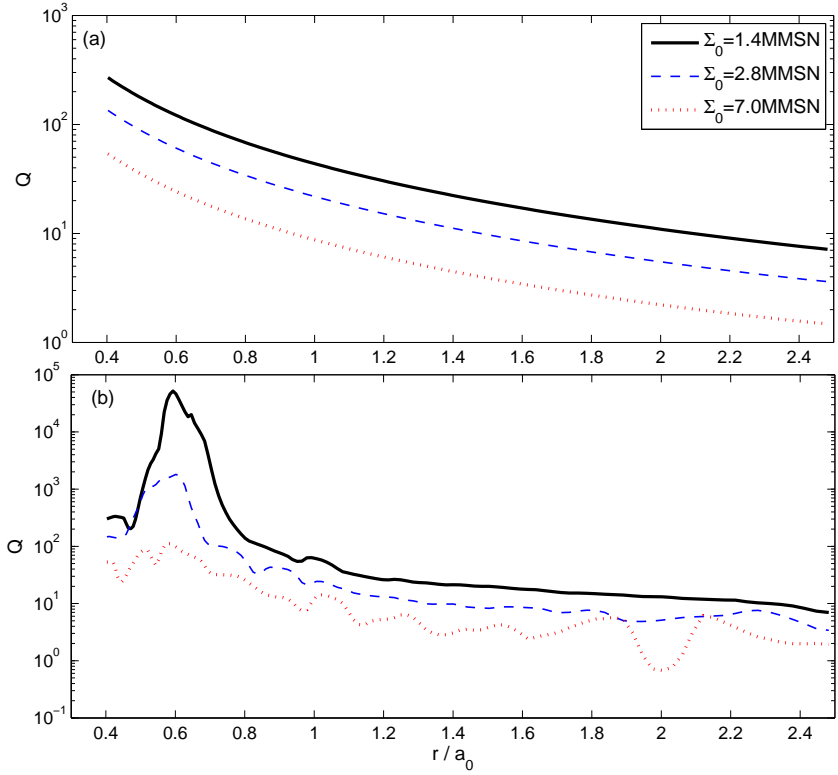}\\
  \caption{$Q$ profile on the disk. Panel (a): The initial $Q$ profiles. Panel (b): The final $Q$ profiles (b) on the disk with different surface density}\label{QvsR}
\end{figure}

To reduce the initial impact on the disk, we hold the planet in a
circular orbit for $50$ orbits and increase its mass from $0.01$ to $1$
Jupiter mass gradually. Since the initial planet mass is very small and the initial velocity of gas has taken the gravity forces and the pressure into account, the disk achieves a steady state well before the planet emerges. Two strong spiral arms emerges after about $30$ orbits. When we release the planet, a clear gap is already formed. At the initial state, Toomre $Q$ is greater than $1$ over the disk(fig. \ref{QvsR}).

The calculations are actually performed in a wide annulus, with the inner
boundary located at $R_{in}=0.4a_0=2.08AU$ and the outer one located at
$R_{out}=2.5a_0=13AU$. We adopt outgoing boundary conditions at both the inner and outer boundaries. It is a wave absorbing boundary condition that the waves are only
allowed to propagate out of the computational domain, while the inward
traveling waves are set to be zero. There are two ghost rings outside
the boundaries, whose density and velocity field stay at the initial
state. In the self-gravitating model, we include the gravity potential
of these two ghost rings to avoid the un-physical cutting-off of the
self-gravity potential at the edges of the disk.

\subsubsection{Measurement of the gap width}
\begin{figure}
    \includegraphics[width=140mm]{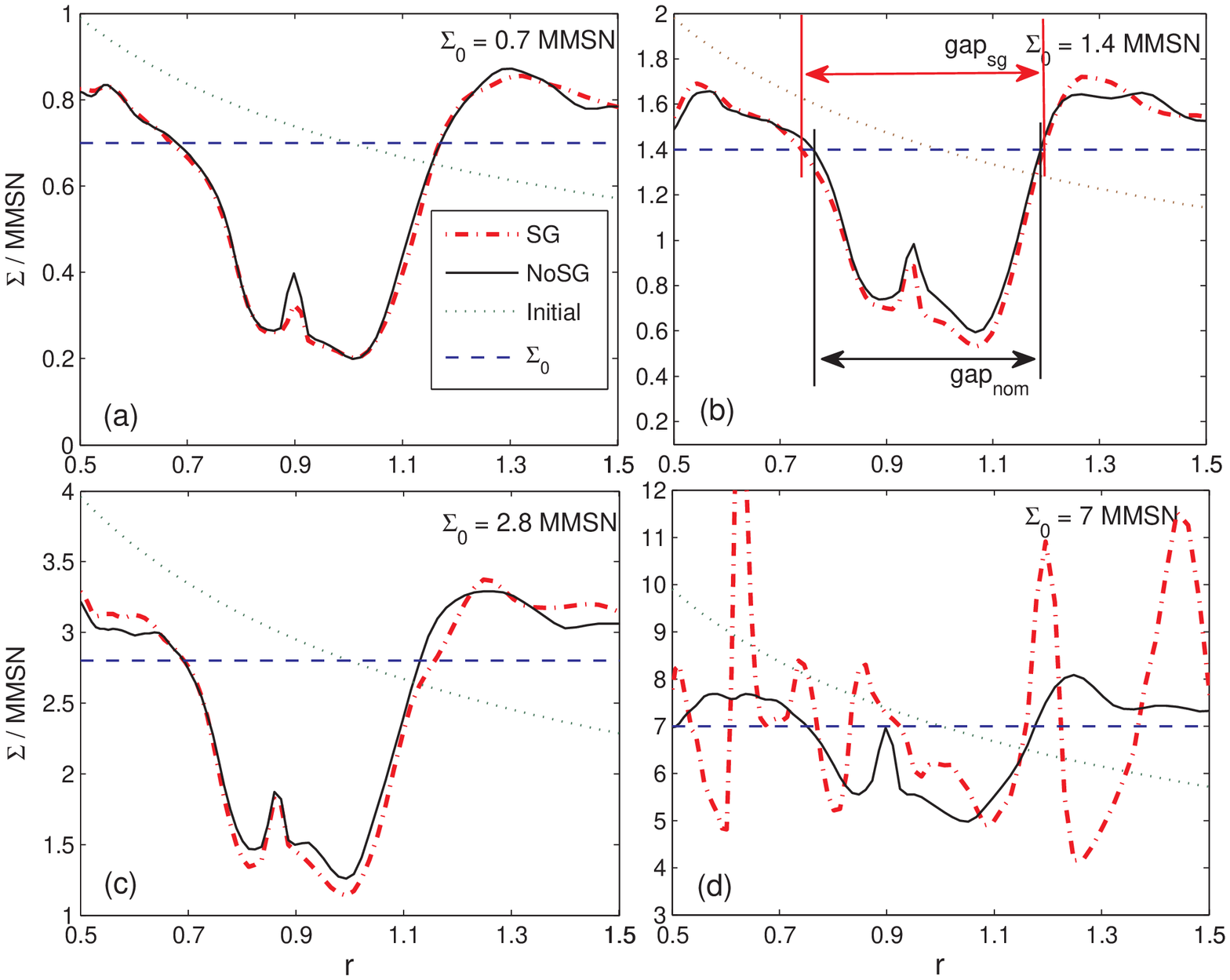}\\
  \caption{Surface density cross sections of the disks with different surface densities. The surface density is averaged over the azimuthal direction. These figures show the gap structures when the disk's self-gravity is included or excluded. When the disk's self-gravity is included, the gap is slightly deeper and wider. Panel (b) shows how we measure the gap width. Panel (d): disk's structure becomes very turbulent in a dense self-gravitating disk. There is no clear gap in that case.}\label{den_cross}
\end{figure}

The gap width is a key quantity in this work, however the exact positions of gap boundaries are hard to be determined analytically. Fortunately, we are focusing on the relative changes of gap width in a disk with or without self-gravitating effects. So, we could define the gap width by the disk's surface density profiles. To ensure the comparability, we set the surface density at the initial position of the planet as the reference density. Then, the measurement of the gap width in 1-D simulation is quite simple. At each side of the planet's orbit, we can find a position where the surface density is equal to the reference density. If we got more than one positions, the nearest one (to the planet) is chosen. Then we get two positions on both sides of the planet. We define this two radii as the inner and outer boundary of the gap and the gap width is the difference of their radial positions. The measurement in 2-D simulation is similar. The only difference is that we use an azimuthal averaged density profile in the 2-D simulations(panel (b) in fig. \ref{den_cross}).

\section{Results}

Our numerical simulations consist of two steps. First, we adopt a 1-D
model that describes the radial viscous evolution of a self-gravitating
disk. The self-gravitating effect of the gas is added in both as an
additional radial force field and an effective viscosity. Since the 1-D
model is not suited to simulate the 2-D gravitational turbulence and the
behavior of a gravitationally unstable disk, we concentrate on a low
surface density range to study the gap variation in a gravitationally
stable disk. Second, to reveal the gap variation within the transition
stage (from gravitationally stable to unstable), we further perform a
series of fully self-consistent 2-D simulations with the self-gravity of
gas included. We then investigate the orbital evolution of the embedded
planet associated with the gap formation process.

\begin{figure}
    \includegraphics[width=140mm]{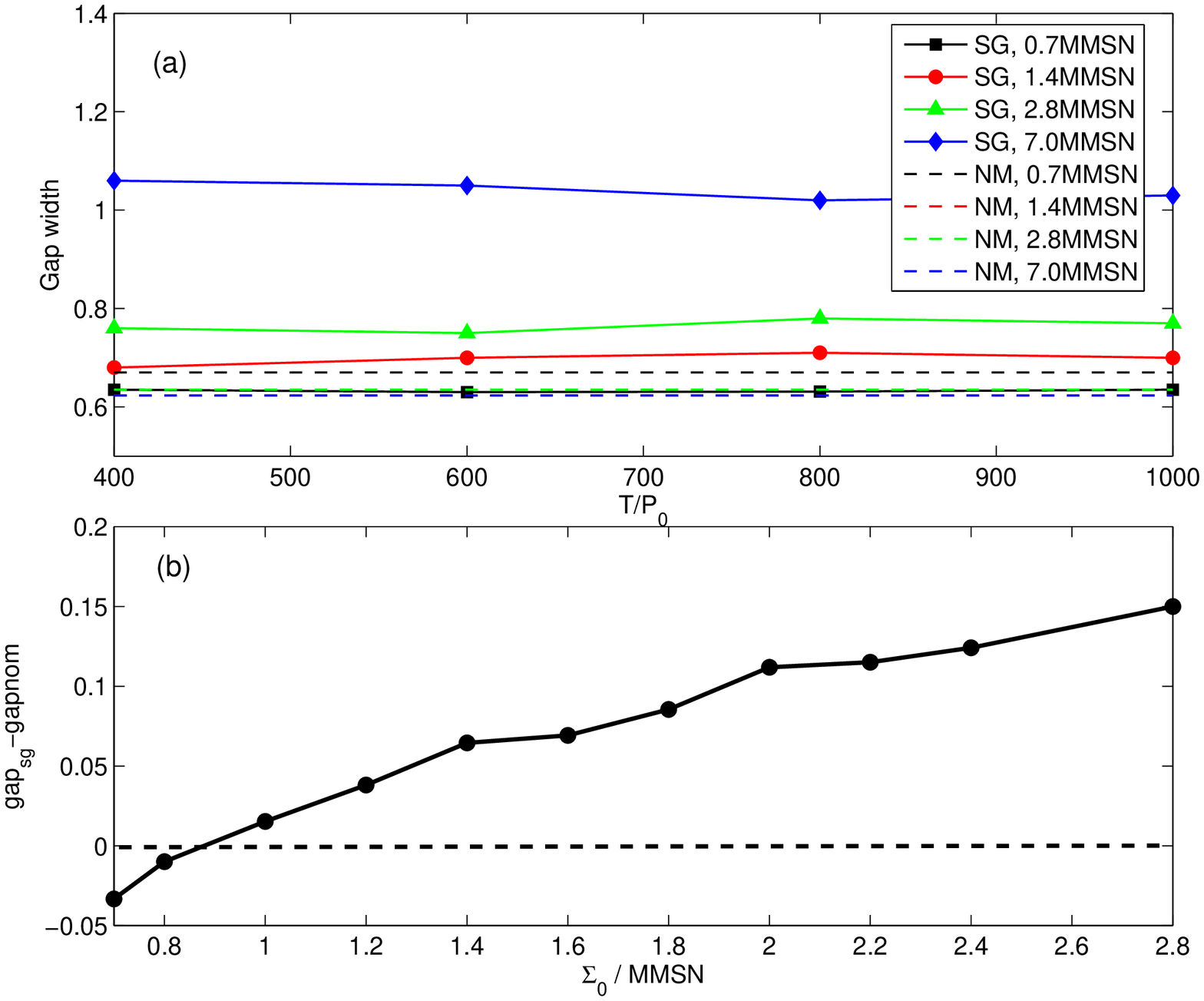}\\
  \caption{Gap widths in 1-D simulations. Panel (a): Width of gap versus evolution time. From top to bottom the surface density of disk decreases from $7MMSN$ to $0.7MMSN$. The dashed lines denote the non-self-gravitating cases. Since the gap size almost doesn't change with disk's density in the non-self-gravitating cases, the dashed lines denote $1.4MMSN$ and $2.8MMSN$ are overlapped. Panel (b): The differences of gap width between the self-gravitating disk and the non-self-gravitating disk. The surface density increases from $0.7MMSN$ to $2.8MMSN$, the critical surface density is around $\Sigma_I\approx0.85MMSN$. }\label{gap_diff_1d}
\end{figure}

\subsection{1-D Simulation}\label{1-D}

Panel (a) of fig. \ref{gap_diff_1d} shows the variation of the gap
width versus evolution time in the self-gravitating and
non-self-gravitating models. Our 1-D simulations show that, in a disk
without the self-gravity effect, the gap width is almost unchanged when
the surface density (or disk mass) increases. This is consistent with
the former analysis that when the self-gravity is absent the gap width
is determined by the dissipation of the gas and the tidal force of the
planet \cite{gol80,lin86}. When the self-gravity is included, we find
the gap width increases as the surface density increases. When the gap
width becomes stable, we measure the width difference between the two
gaps in different models for a series of surface densities (panel (b) of
fig. \ref{gap_diff_1d}). It clearly shows that there exists a critical
surface density around $\Sigma_I \simeq 0.85MMSN$. The self-gravity
suppresses the gap formation process when $\Sigma_0 < \Sigma_I$ and
enlarges the gap when $\Sigma_0 > \Sigma_I$.

During the gap formation process, the self-gravity effect plays two
opposing roles. On one hand, it drives an effective viscosity
\cite{gam01} which tends to make the disk more dissipative. Therefore,
the gap is more difficult to be cleared and the gap formation process is
suppressed. On the other hand, the equilibrium at the position of the
gap boundaries changes as the local self-gravitational potential varies
with the surface density there. When the density slope becomes sharp at
the gap boundaries, the local self-gravity potential may change
direction and tends to contract the disk. This effect may leads to
enlargement of the gap. The behavior of the gap width under this two
effects is described below.

When the disk surface density is low, the dynamics of the gas are mostly
determined by the central gravity $GM_\ast/r^2$. Although the global
self-gravity potential of the disk is weak, the gas exchanges angular
momentum more effectively with immediate neighbors by the local mutual
gravity. This can be expressed as an effective viscosity which
suppresses the gap formation process. As the disk density increases, the
global self-gravitational potential begins to make measurable influences
to the central gravity. We could just look at the outer boundary of the
gap, where $r=r_{ob}$. When the gap is stable, there is an equilibrium:
\begin{equation}
\frac{v_\theta^2}{r_{ob}}=\frac{GM_\ast}{r_{ob}^2}+\frac{1}{\Sigma}\frac{dP}{dr}|_{ob},
\end{equation}
given that the tidal force of the planet is balanced by the viscosity
dissipation. When the self-gravity $\Phi_{sg}(r,t)|_{ob}$ is included,
the equilibrium becomes:
\begin{equation}
\frac{v_\theta^2}{r_{ob}}=\frac{GM_\ast}{r_{ob}^2}+\frac{1}{\Sigma}\frac{dP}{dr}|_{ob}+\frac{d\Phi_{sg}}{dr}|_{ob}.
\end{equation}

As the gas is being cleared in the gap, the gradient of
self-gravitational potential becomes very sharp at the boundaries. Thus,
$F_{sg}(r_{ob},t)=-d\Phi_{sg}/dr|_{ob}$ increases from negative (directs
inward) to positive (directs outward). By assuming the tidal force of
the planet and the viscous dissipation remain balanced, we may find that
when $F_{sg}(r_{ob},t)$ increases, the angular velocity required by the
equilibrium decreases. During this transition stage, the angular
velocity of the gas at $r_{ob}$ is greater than that required by the
equilibrium. So the gas tends to drift outward. At the meanwhile, the pressure
gradient and viscous dissipation try to push the gas back. However, the disk has not been dominated by the gravitational turbulence yet---the effective viscosity is still too low: $\alpha_{sg}\sim 10^{-3}$. The viscous timescale
is as long as $10^{6}yr$ and is much longer than the variation time
scale of $F_{sg}(r_{ob},t)$ which is only dozens of orbits for a Jupiter
mass planet. To retain the equilibrium, the surface density profile
needs to become sharper to generate a stronger pressure gradient,
$(1/{\Sigma})dP/dr|_{ob}$, at the gap boundaries. However, the sharper
gradient of the surface density also enhances the gradiant of the self-gravity potential at the boundaries. Finally, the outer boundary moves outward until the
angular velocity of the gas matches the required value and a new
equilibrium is achieved. A similar process occurs at the inner boundary
of the gap but results in an inward drift of the gas. This combined
effect behaves like a 'self-gravitational contraction' of the two parts
of the disk and makes the gap become wider and deeper. Furthermore,
since the pressure effect decreases as $\Sigma$ increases, this effect
is more pronounced as the disk becomes denser
(Fig.\ref{gap_diff_1d}).

\subsection{2-D Simulation}

Our 1-D simulations suggest that when the surface density exceeds
$\Sigma_I$, the width of the gap increases monotonically (for $\Sigma$
up to $2.8MMSN$). To ensure this trend in a fully described
self-gravitating disk, a series of 2-D hydrodynamic simulations are
performed. The orbital evolution of the giant planet embedded in a
self-gravitating disk is also studied. Since the 2-D simulation is very
time consuming when the disk self-gravity is included, we chose only 4
typical surface densities: $0.7MMSN$, $1.4MMSN$, $2.8MMSN$ and $7MMSN$.

\begin{figure}
    \includegraphics[width=140mm]{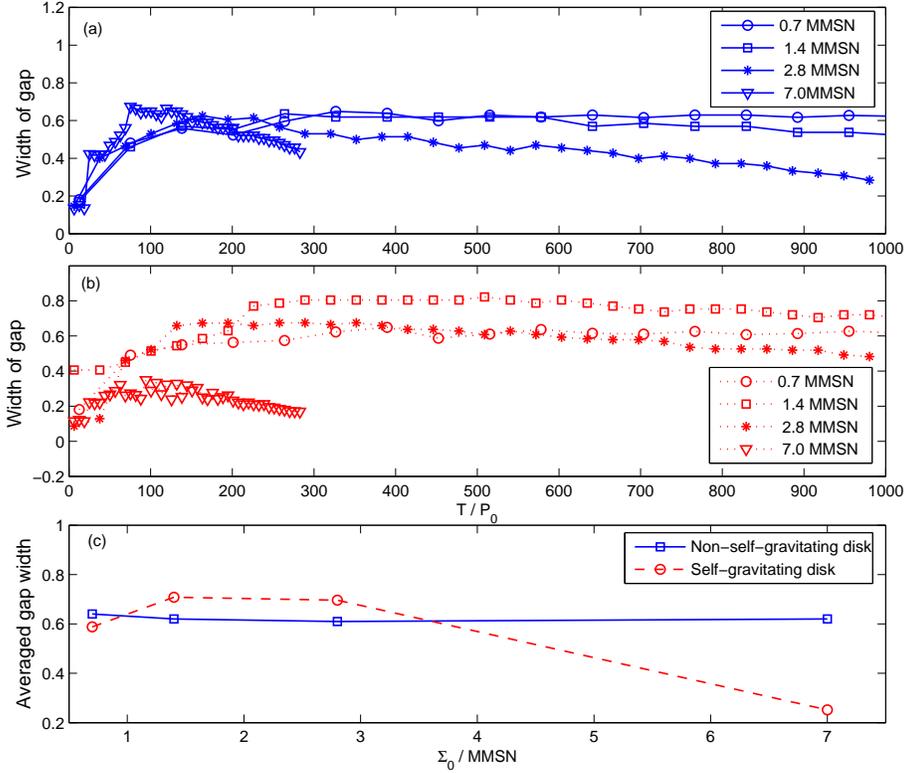}\\ 
  \caption{The evolution of gap width in 2-D simulations.
  Panel (a): gap width versus evolution time in non-self-gravitating disks. The gap is cleared around $100$ orbits and reaches the maximum value. Then, the gap width decreases as the planet migrates inward. All the widths are azimuthal averaged.
Panel (b): gap width versus evolution time in self-gravitating disks.
Panel (c): differences of gap width in self-gravitating and non-self-gravitating
cases. The gap width does not change with the surface density of the
non-self-gravitating disks. }\label{gap_width_2d}
\end{figure}

\subsubsection{Gap formation}\label{gap formation}

\begin{figure}
    \includegraphics[width=140mm]{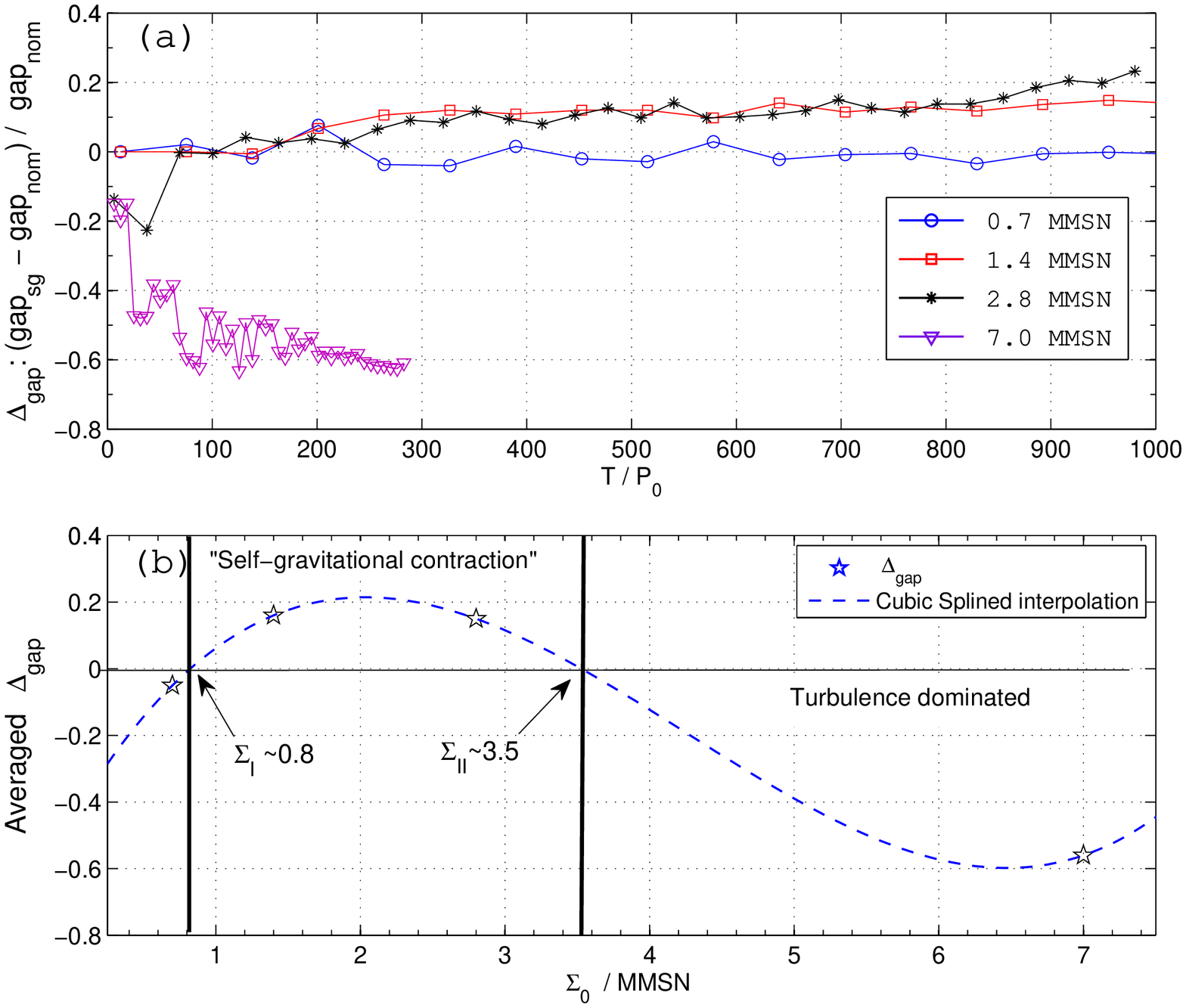}\\
   \caption{The relative differences of gap width in 2-D simulations. Panel (a):
The relative width differences between self-gravitating and non-self-gravitating
disk. Gap width has been normalized by the corresponding semi-major axis of the planet first to eliminate the migration effect.
Panel (b): The time averaged relative differences versus surface density of disk. The
dashed lines show the cubic spline interpolations and we found the
second critical surface density $\Sigma_{II}\approx3.5MMSN$. The first
one $\Sigma_{I}\approx0.8 MMSN$ is coherent with our 1-D results. When
$\Sigma_{II}>\Sigma_0>\Sigma_I$ the ``self-gravitating contraction''
dominates the gap formation process. When $\Sigma_0>\Sigma_{II}$,
the gravitational turbulence viscosity becomes dominant. In a self-gravitating disk, gap width reaches maximum when $\Sigma_0\approx 2MMSN$. All the widths we adopt are the azimuthal averaged value. Notice that, for a very dense disk $\Sigma_0\geq 7MMSN$, the gap is not cleared. So it is significantly small. }\label{gap_diff_2d}
\end{figure}

\begin{figure}
    \includegraphics[width=180mm]{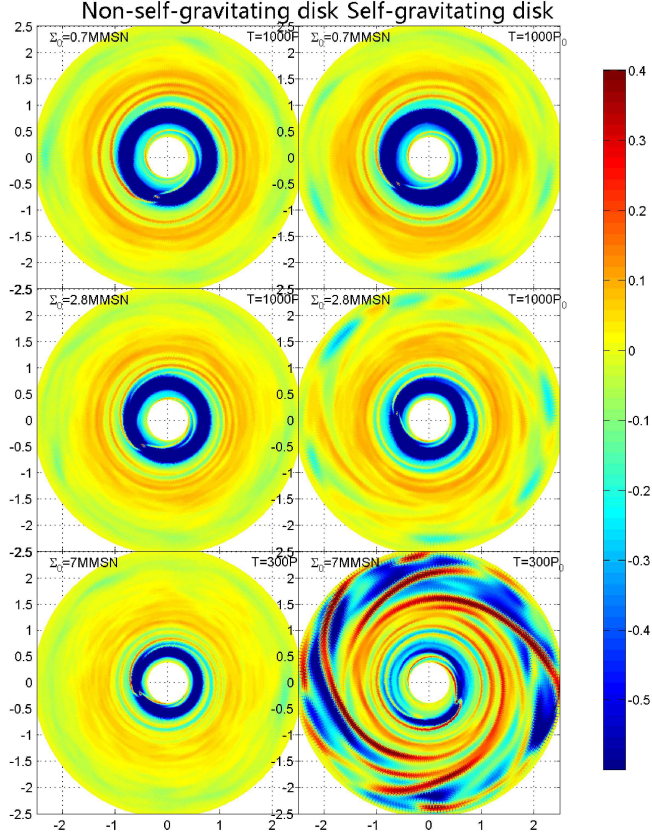}\\
  \caption{The surface density evolution. The left panels
show the non-self-gravitating disk and the right panels show the
self-gravitating disk. The surface density of disk is $\Sigma_0=1.4MMSN,
2.8MMSN, 7MMSN$ from top to the bottom, respectively. The gravitational
turbulence emerges clearly when $\Sigma_0>2.8MMSN$. }\label{density6}
\end{figure}

Panel (a) and (b) of fig. \ref{gap_width_2d} shows the evolution of gap width
versus time in the self-gravitating and non-self-gravitating disks.
Different surface densities are denoted by corresponding marks. Note
that the decrease of the gap width after about $200P_0$ is due to the
decrease of the Hill radii when the planet is migrating inward($a_p$
decreases). It is clear that the surface density does not change the gap width when the self-gravity is excluded, while the gap width in a self-gravitating
disk strongly depends on the surface density (see panel (c) of Fig.
\ref{gap_width_2d}). Panel (a) of fig. \ref{gap_diff_2d} shows the evolution of the normalized gap differences: ($gap_{sg}-gap_{nom})/gap_{nom}$. All the gap widths have been normalized by the corresponding semi-major axis of the planet to eliminate the migration effect. In a self-gravitating disk, the gap width increases as the disk's surface density increases. However, it is not a linear relation.
Furthermore, the 2-D simulations show that, the enlargement of the gap decreases when the surface density exceeds $\sim 2MMSN$ and becomes negative when $\Sigma_0>3.5 MMSN$(panel (b) of fig. \ref{gap_diff_2d}). The gap width is recorded every $10$ orbits, when the simulation is finished, we sum all the widths together and find the averaged value. Note that gap widths of the first $100$ orbits are dropped, since the gap is not well formed before that. fig. \ref{den_cross} shows the gap structures under different situations and how we measure the gap width. The gap size is almost identical when the surface density is low. When the disk becomes denser, the gap is slightly deeper and wider in the self-gravitating disks. We measure the differences of the averaged gap width between the self-gravitating and non-self-gravitating models and interpolate these data (panel (b) of fig. \ref{gap_diff_2d}). The results suggest that there is another critical surface density which is around $\Sigma_{II}\simeq3.5MMSN$. When $\Sigma_0\geq \Sigma_{II}$, the self-gravity suppresses the gap formation again. Notice that, for a very dense disk $\Sigma_0\geq 7MMSN$, the gap is not cleared. So it is significantly smaller than others(Panel (d) of fig. \ref{den_cross} and fig. \ref{density6}). The widths showed in fig. \ref{gap_width_2d} are azimuthal averaged value.

When the surface density exceeds $2.8MMSN$, the gravitational turbulence
becomes significant. fig. \ref{density6} show the density contours of the disk under different surface densities. The three figures in the left column are the normal disks. Their disk structure do not change much when their surface density increases from $0.7MMSN$ to $7MMSN$. The three figures in the right column are the self-gravitating disks. When the surface density increases to $2.8MMSN$, the turbulence emerges at
the outer part of the disk, where the Toomre $Q$ is relatively low. As
the surface density increases more, the gravitational turbulence becomes
stronger. When the disk's surface density exceeds $\Sigma_{II}$, the disk becomes gravitational unstable(the last figure in the left column).  At such high surface density, the effective viscosity caused by the self-gravitational turbulence will overcome the 'self-gravitational contraction' effect and dominate the gap formation process.

Comparing with our 1-D simulations, there are two major differences. One
is that our 2-D simulations indicate a smaller value of the first
critical surface density $\Sigma_I \sim 0.8$. This suggests that the
'self-gravitational contraction' is stronger in a 2-D disk. It is
probably because in the 1-D simulations, we adopt an artificial
viscosity $\nu_{art}$, which turns out to be slightly larger than the
numerical viscosity $\nu_{num}$ in our 2-D simulations. This makes the
total effective viscosity in the 1-D simulation slightly larger than the
one in the 2-D simulation. However, the difference is quite small (our
1-D results indicate $\Sigma_I \sim 0.85$) and doesn't change our main
results.

The other difference is that our 1-D results suggest that the gap size
increases monotonically as the surface density increases from $0.7MMSN$
to $2.8MMSN$. However our 2-D results show that the increasing trend
bends down around $2MMSN$. For the 1-D simulations, the angular momentum
exchange caused by the self-gravity effect was only described by an
effective viscosity $\nu_{sg}$. In this description,
$\nu_{sg}=\alpha_{sg} c_s^2 / \Omega$(\cite{gam01}), where
\begin{equation}
\alpha_{sg}=\frac{4}{9\gamma(\gamma-1)t_{cool}\Omega}.
\end{equation}
The cooling time scale is determined by the internal energy per unit area $U$ and the cooling function $\Lambda$,
\begin{equation}
  t_{cool}=\frac{U}{\Lambda}=\frac{c^2_s\Sigma}{\gamma(\gamma-1)\Lambda},
\end{equation}
and(\cite{hub90})
\begin{equation}
  \Lambda=\frac{16\sigma}{3}(T^4_c-T^4_o)\frac{\tau}{1+\tau^2}.
\end{equation}
$T_c = 280K(a/1AU)^{-1/2}$ is the mid-plane temperature of the disk and
$T_o=10K$ is a minimum temperature of background sources(\cite{sta07}).
Using the analytic approximation of the Rosseland mean opacity for
molecules (\cite{bel94}),
\begin{equation}
  \kappa=\kappa_0(\frac{\Sigma}{2H})^{2/3} T^3_c,
\end{equation}
we can get the optical depth $\tau$ (\cite{ric10}),
\begin{equation}
  \tau\approx H\kappa(\frac{\Sigma}{2H},T)\frac{\Sigma}{2H}=H(\frac{\Sigma}{2H})^{5/3} T^3_c.
\end{equation}
Then we get
\begin{equation}
  \Lambda=\frac{16\sigma}{3}(T^4_c-T^4_o)\frac{H(\frac{\Sigma}{2H})^{5/3} T^3_c}{1+(H(\frac{\Sigma}{2H})^{5/3} T^3_c)^2} .
\end{equation}
At the location of the giant planet, where $a=5.2AU$, $H/r=0.02$ and $\Sigma=1MMSN$, we found that
\begin{equation}
(H(\frac{\Sigma}{2H})^{5/3}T^3_c)^2 \approx 10^3 \gg 1.
\end{equation}
Thus, we have $\Lambda \propto \Sigma^{-5/3}$. This gives us $t_{cool}
\propto \Sigma^{8/3}$ and $ \nu_{sg} \propto \Sigma^{-8/3}$. So, as the
surface density $\Sigma$ increases, the dissipation in the disk becomes
weaker and the gap forms more effectively. This result could be valid
when $Q$ is much larger than unity (\cite{ric10} estimated that $Q \geq 2$).
In some high-density simulations, however, $Q$ is close to unity after several
hundred orbits (fig. \ref{QvsR}). So we believe that the real self-gravitating effect of a dense disk should be calculated by the realtime density distribution consistently and the 2-D simulations should be more self-consistent.

\begin{figure}
    \includegraphics[width=140mm]{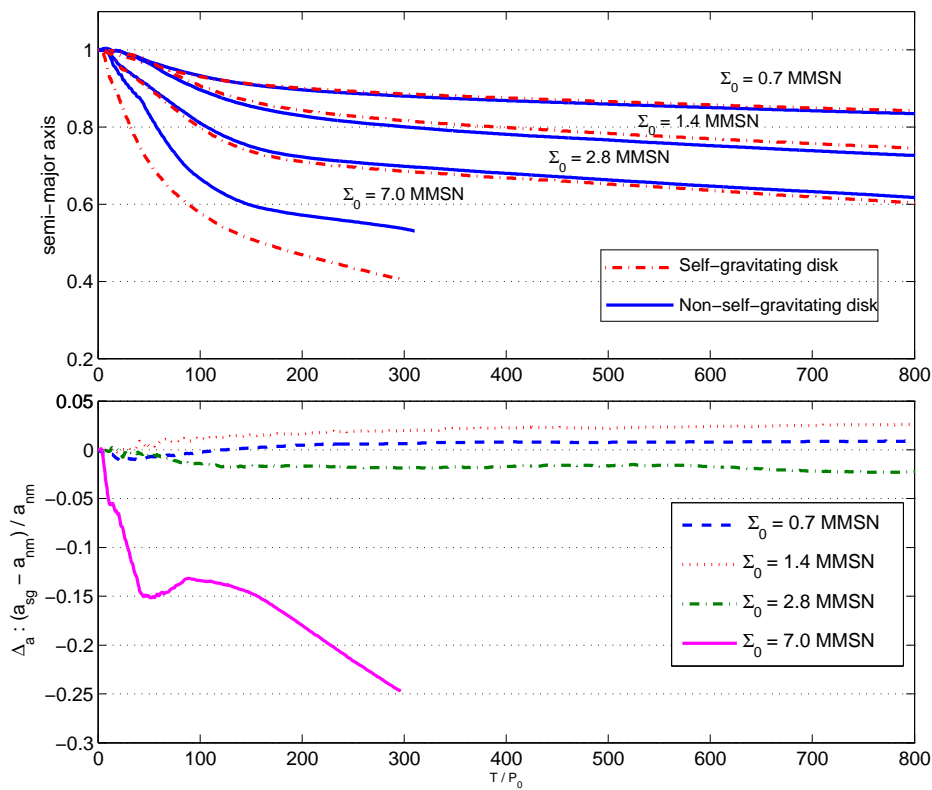}\\
  \caption{Upper panel: orbital migrations of the planet with or without disk's self-gravity. Blue solid lines show the migration in non-self-gravitating disks, while the red solid lines show the migration in self-gravitating disks.
  Lower panel: the differences of migrations between the self-gravitating cases and non-self-gravitating cases(normal cases). The value has been normalized by the value from corresponding normal cases. The difference on the migration is significant only when the disk is very dense.}\label{at4}
\end{figure}

\subsubsection{Migration of the giant planet}\label{migration}
\begin{figure}
    \includegraphics[width=140mm]{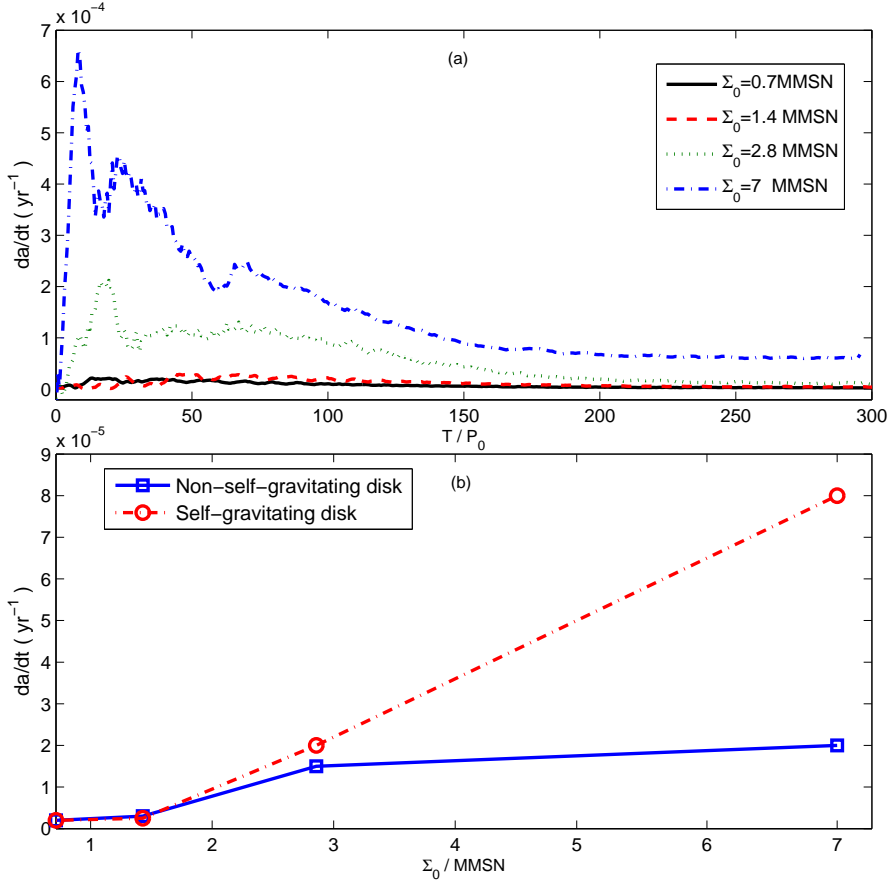}\\
  \caption{The migration rate $\dot{a}$ versus surface density of disk.
Panel (a): The migration rate versus evolution time in the
self-gravitating disk with varies surface densities. $\dot{a}$ reaches
steady value after the gap is cleared ($t=200$ orbits). Panel (b): The
migration rate versus surface density of disk. In the self-gravitating
disk, $\dot{a}$ is proportional to $\Sigma_0$ (red circles). While in
the non-self-gravitating disk, this relation is very weak (blue
rectangles).  }\label{dadt}
\end{figure}

Besides the gap formation, the orbital migration of the planet is
another important outcome of the disk-planet interactions. The upper panel of fig.
\ref{at4} shows the migration of the planet embedded in a series of
disks. The dashed lines are the results with the self-gravity of the gas
included, while the solid lines are those results without the
self-gravity of gas. From top to the bottom, the surface density of the
disk increases from $0.7MMSN$ to $7MMSN$. One may find that all the
migrations experience two stages. At the first stage, the giant planet
is still surrounded by the gas and undergoes the type I (or type-I-like)
migration whose time scale should be inversely proportional to the
disk's surface density (\cite{tan02}),
\begin{equation}
\tau=(2.7+1.1\gamma)^{-1}\frac{M_\ast}{M_p} \frac{M_\ast}{\Sigma_0 a^2_p}(\frac{c_s}{a_p \Omega_p})^2\Omega^{-1}_p \propto \Sigma^{-1}_p
\end{equation}
Our results show that at this stage, the migration rates of the planet
is greater as the disk becomes denser(slope of the migration curve in the upper panel of fig.\ref{at4} and panel (a) in fig. \ref{dadt}). It is qualitatively agree with the analytic predictions we mentioned above and this could be a proof of the consistency of our simulations. The lower panel of fig.\ref{at4} shows the relative differences of migration (semi-major axis vs. time) between the self-gravitating cases and the normal (non-self-gravitating) cases. The differences are normalized by the values from the corresponding normal cases.

As the gas located in the gap region is cleared, the migration of the
giant planet steps into the second stage when the migration rate of the
planet is significantly reduced. This is usually called type II
migration. According to linear analysis, the time scale of type II
migration is supposed to be inversely proportional to the effective
viscosity on the disk. From fig. \ref{at4} we can find that the type II migrations in different surface density have almost the same slope when the self-gravity of disk is exclude. This is reasonable since the effective viscosity should not depend on the surface density. However, we find that the migration rate in the denser disk is indeed larger than the rate in the thinner disk(also in panel (b) of fig.\ref{dadt}). The reason is that there is an inner boundary in our disk model. When the planet is getting close to the inner boundary, most of the inner disk has flow outside our inner boundary. As a result, the torque from the inner disk(positive torque) is weaken and the net negative torque is greater. That means the planet will drop to the central star faster when it gets closer to the inner boundary in our simulations. At the meanwhile, a planet migrates faster in a denser disk than in a thinner disk before the gap is cleared. So, when the migration steps into the type-II regime, a planet embedded in a denser disk will be closer to the inner boundary of the disk and will has larger inward migration rate.   While in the self-gravitating disk, the type II migration rate changes versus the disk's surface density, because the effective viscosity is related to the disk's surface density now. When the surface density is low, the difference of the semi-major evolution is very small: $<2\%$ (lower panel of fig.\ref{at4}). When the surface density is higher($\Sigma_0\geq7.0MMSN$), the difference becomes very significant.

In this paper, we concentrate on the variations
of the migration rate under the effect of disk's self-gravity which is
the source of the turbulent viscosity. Panel (a) of fig. \ref{dadt}
shows the evolution of the migration rate ($\dot{a_p}$) with the
self-gravitating effect included. After about $300$ orbits, the
migration rate reaches different stable values according to the surface
density of the disk. We measure this stable migration rate in each run
and the results are shown in panel (b) of fig. \ref{dadt}. The red
circles are the results with self-gravity of gas included and the blue
squares are the results without the self-gravity. The migration rate
weakly increases with the surface density in a non-self-gravitating
disk. This indicates that the effective viscosity barely changes with
the surface density when the self-gravitating effect is excluded.
However, in a self-gravitating disk, the migration rate increases
quickly as the surface density increases. Our results suggest that, in a
self-gravitating disk, the migration of a giant planet is slightly
slowed (almost identical with the non-self-gravitating case) when the surface density is moderate. However, the migration of the giant planet becomes faster than that in the
non-self-gravitating disk when the surface density exceeds $2.8MMSN$. In
a very dense disk $\Sigma_0=7MMSN$, the migration of the giant planet
could be very fast and the time scale could be as short as $\sim10^4 yr$
(fig. \ref{dadt}).

\begin{figure}
    \includegraphics[width=140mm]{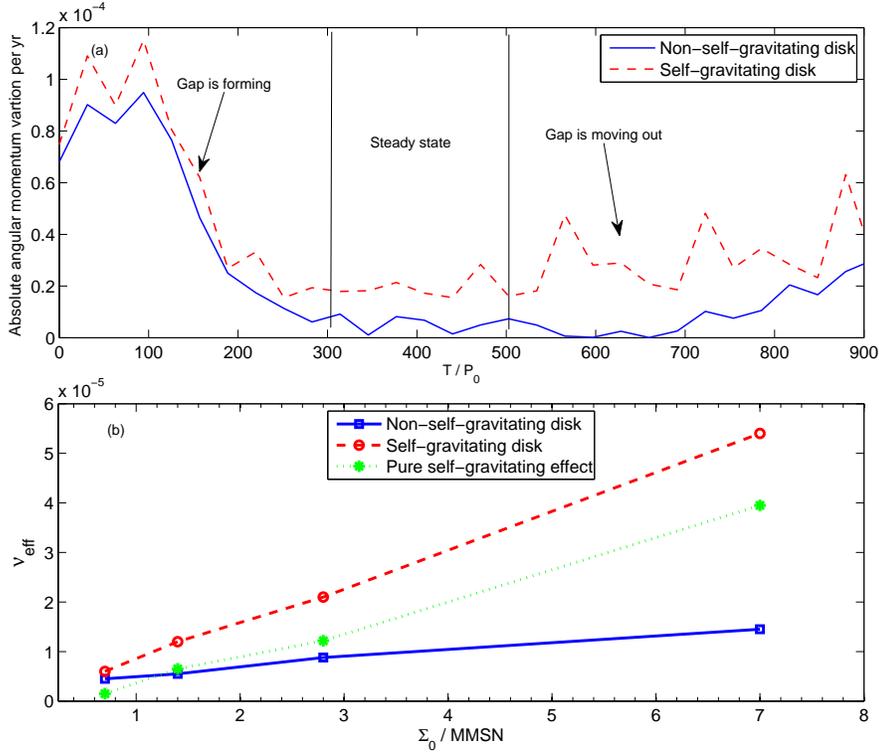}\\
  \caption{Panel (a): The absolute angular momentum variations versus evolution time. In a self-gravitating disk, the angular momentum variation (red dashed line) is always larger than that in a non-self-gravitating disk (blue solid line), where $\Sigma_0=2.8MMSN$. Panel (b): The effective viscosity versus surface density of disk. In a non-self-gravitating disk, the effective viscosity is mainly due to the numerical dissipation $\nu_{eff}=\nu_{num}$ (blue rectangles). While in a self-gravitating disk, the gravitational turbulence is the main source of dissipation $\nu_{eff}=\nu_{num}+\nu_{sg}$ (red circles). The net gravitational viscosity effect $\nu_{sg}$ is then shown by green stars, which is proportional to $\Sigma_0$. }\label{nu_eff}
\end{figure}

The quick increase of the migration rate indicates that the effective
viscosity is mostly determined by the gravitational turbulent viscosity
and increases with the surface density of a self-gravitating disk. We sum all the angular momentum of the whole disk and measure its variation rate. Since the
size of the disk does not change with time, its angular momentum
variation is only determined by the radial mass flow and angular
velocity variation, which are both the results of the viscous
dissipation when the gap is stable. Therefore, this angular momentum
variation rate will roughly indicate the effective viscosity $\nu_{eff}$
on the disk. The associated results are shown in fig. \ref{nu_eff}.
Panel (a) shows the angular momentum variations versus time in a disk
where $\Sigma_0=2.8MMSN$. The result with the disk's self-gravity
included is denoted by the red dashed line, and the one with
self-gravity excluded is denoted by the blue solid line. The large
variation rate before $t=300P_0$ is the result of the gap formation
process, where gas is driven away by the tidal torque of the planet and
results in a sharp decrease in the total mass of the disk. When the
planet migrates significantly ($t>500P_0$), the gap moves close to the
inner boundary of the disk. The total angular momentum of the disk
increases as the gap moves out of the disk's inner boundary (total mass
increases). We estimate the averaged dissipation rate only at the steady
state of each run ($300P_0<t<500P_0$) and the results are shown in panel
(b) of fig. \ref{nu_eff}. Since we do not adopt any artificial
viscosity, for a non-self-gravitating disk $\nu_{eff}=\nu_{num}$, and
for a self-gravitating disk $\nu_{eff}=\nu_{num}+\nu_{sg}$. Our results
show that the effective viscosity $\nu_{eff}$ increases with $\Sigma_0$
in the self-gravitating disk. For the non-self-gravitating disk, the
$\nu_{eff}$ only slightly increases with $\Sigma_0$. Then we find that
$\nu_{sg}$ is roughly proportional to $\Sigma_0$ (green stars in panel
(b) of fig. \ref{nu_eff}).

\begin{figure}
    \includegraphics[width=140mm]{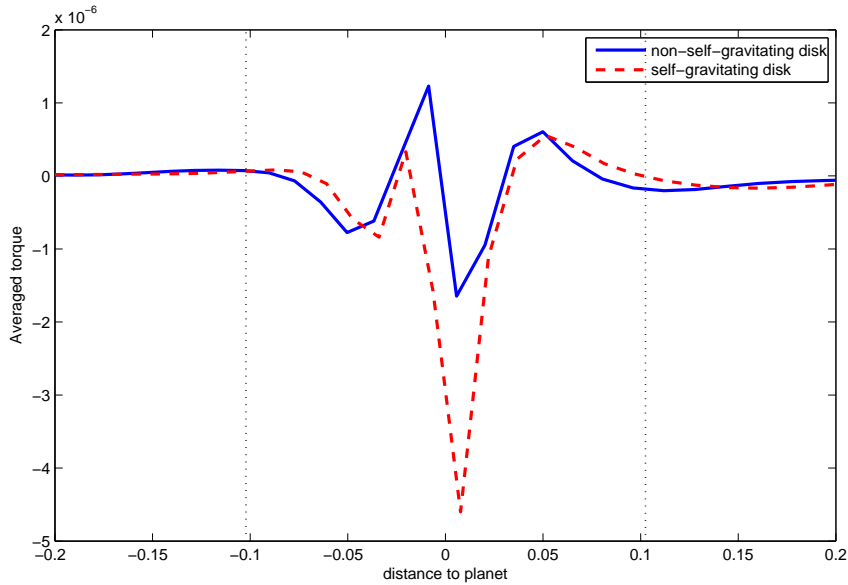}\\
  \caption{Azimuthal averaged torque in the vicinity of the giant planet embedded in a very dense disk. The surface density of disk is $7MMSN$. The torque is almost symmetric with planet's position in the non-self-gravitating disk(blue solid line). While in a self-gravitating disk, the planet suffers a net negative corotation torque(red dashed line).}\label{averaged_torque}
\end{figure}

These results are in very good quantitative agreement with the migration
rates we obtained above, except for the very high surface density
$\Sigma_0=7MMSN$, where the migration time scale ($\sim1.2\times10^4
yr$) is much shorter than the viscous time scale ($\sim2.1\times10^4
yr$). In fact, in such a dense disk, the planet cannot clear a gap
before it reaches the inner boundary (fig. \ref{density6}). As the
Toomre $Q$ decreases along the disk radius, the gravitational turbulence
becomes stronger as the radius increases. This generates a vortensity
gradient across the corotation region of the giant planet and exerts a
large negative corotation torque on the planet \cite{mas03}. We further
calculate the torques exerted on the planet. Figure
\ref{averaged_torque} shows the azimuthal averaged torque as a function
of the distance to planet. It clearly shows that, in a
non-self-gravitating disk, the torque density is almost symmetric with
the position of planet. There is a great negative torque within the
corotation region of the planet, which drags the planet inward even
faster. This result is in good agreement with that obtained by Baruteau
\textit{et al.} (2011).

\section{Conclusions and Discussions}

In this paper, we concentrate on the gap formation process under the
effect of a disk's self-gravity. We first perform a series of 1-D
simulations, where the disk's self-gravity is modeled by a gravitational
effective viscosity $\nu_{sg}$ and a time dependent azimuthal-averaged
self-gravity potential. We find that when the surface density of the
disk is low, the self-gravity potential is too weak to affect the gap
formation process and the gravitational effective viscosity suppresses
the growth of the gap. As we increase the surface density of the disk,
the self-gravitational potential becomes stronger. It leads to a
'self-gravitational contraction' effect at each boundary of the gap and
tends to enlarge the gap size. When the surface density exceeds a
critical value, $\Sigma_0>\Sigma_I$, the net self-gravitating effect
begins to benefit the gap formation process and the gap width increases
with the surface density of the disk. We estimate this critical surface
density is around $\Sigma_I\approx0.8MMSN$ (section \ref{1-D}). Since we
recognize that the gravitational turbulence viscosity could not be
described consistently in a 1-D simulation, we further perform a series
of 2-D simulations where the disk's self-gravity is fully calculated by
the real-time density distribution on the disk. We find that the width
of the gap will not increases with the surface density monotonically in
a self-gravitating disk. The gravitational turbulence becomes stronger
as the disk's surface density increases and the associated effective
viscosity overwhelms the 'self-gravitational contraction' effect when
the surface density of disk exceeds another critical value
$\Sigma_{II}$. We estimate $\Sigma_{II}\approx 3.5MMSN$ (section\ref{gap
formation}). The value of $\Sigma_{I}$ and $\Sigma_{II}$ depend on the disk settings. Here we only gives the typical ones. Especially for $\Sigma_{II}$, to find its exact value more surface densities beyond $2.8MMSN$ are needed to be tested.

The associated migration rate of the giant planet is also studied in
this paper. Our 2-D simulations show that the migration rate of the
giant planet is slightly reduced in a self-gravitating disk with moderate surface
density ($\Sigma_0<2MMSN$, see fig.\ref{at4}). However, it increases with
the surface density of the disk where the gravitational turbulence
becomes dominant. When the planet is still able to open a clear gap on
the disk, its migration rate is just proportional to the effective
viscosity due to the gravitational turbulence. Furthermore, in a very
dense disk $\Sigma_0>7MMSN$, the strong effective viscosity prevents the
gap formation even for a Jupiter mass planet. The migration timescale
then becomes much shorter than the viscous timescale $\sim 10^{4} yr$.
This is caused by a large negative corotation drag which is the result
of the vortensity gradient around the planet (section\ref{migration}).

According to our results we find that, (1) the self-gravitating effect
may not be treated as simply an effective viscosity, especially for a
moderate surface density. Our simulations reveal that the self-gravity
plays two opposite roles in the gap formation process at the same time
and the net effect depends on the surface density of the disk. (2) The
gravitational viscosity and the associated migration rate of the giant
planet increase with the surface density in a dense self-gravitating
disk ($\Sigma_0>2.8MMSN$). For a very dense disk $\Sigma_0\geq7MMSN$,
where giant planets usually form, the the gravitational effective
viscosity is too strong to allow a clear gap to form and the migration
timescale of a giant planet could be much shorter than the type II
migration.

So, a giant planet is unlikely to stay at large separation from the
central star if the disk is still dense after the planet has formed.
This is not a problem for the core accretion model. A planet core
usually needs $10^{6-7}yr$ to reach $10M_e$(\cite{miz80}), while the gas
disk would been dispersed within $10^6 yr$ (\cite{wol96}). If the giant
planet could successfully form, its migration would be very slow or even
be stopped since the disk is already too thin to generate large
gravitational viscosity and could not deliver enough angular momentum
effectively. The problem is, because of the long timescale required by
the core growth stage, a giant planet is unlikely to form in a wide
orbit by the core accretion model(\cite{dod09}). For a multiple-planet
system, if the outer planet is smaller than the inner one, the two
inward migrating planets may become trapped into mutual mean motion
resonance and migrate outward together(\cite{Zha10a};\cite{Zha10b}). This could
be an effective way to form giant planets at large separation from their
host star. For a single giant planet, however, it is still a problem.
Some work shows that the radiative effect may affect the direction of
the migration and could result in outward migration(\cite{kal08};\cite{bit10}).

If the giant planet forms through the gravitational fragmentation of a
very dense disk, it would probably migrate inward quickly. However, we
emphasize that we do not adopt any cooling process in our 2-D
simulations. This is because we do not want to introduce any poorly
understood factors in our simulations, which would add too many
uncertainties to the results. In our 2-D simulations, we assume a very
cold disk with $H/r$ fixed at $0.02$ and adopt a locally isothermal
equation of state, therefore the cooling in our model is in fact
perfect. Hence, the effective viscosity due to the self-gravitational
turbulence increases with the surface density of the disk and results in
fast inward migration in a dense disk. If proper cooling process were
included, the gravitational viscosity would become less effective,
slowing the migration rate of the giant planet. This should be fully
considered in future work.

We also notice that the existence of the giant planet may trigger the
onset of gravitational instability in the disk. Strong spiral structures
caused by the giant planet may generate a local minimum of $Q$ and cause
global instability when the averaged $Q$ is still far above unity
(fig.\ref{QvsR} and fig.\ref{density6}). This effect depends on the mass
of the giant planet and the disk where it is embedded. The details are
also the subject of future work under preparation.

\begin{acknowledgements}

This work is supported by National Natural Science Funds for Young Scholar(No. 11003010), National Natural Science Foundations of China (Nos. 10833001,10925313 and 11078001), the Research Fund for the Doctoral Program of Higher Education of China (Nos.20090091110002 and 20090091120025). We are grateful to the High Performance Computing Center (HPCC) of Nanjing University for doing the numerical calculations in this paper on its IBM Blade cluster system. RW is supported by an UNSW Vice-Chancellor's Fellowship.
\end{acknowledgements}

{}

\appendix

\section{Refined treatment of gravity in the vicinity of planet}\label{torque}

When we calculate the torque exerted on the planet by a single cell of
gas, the mass of this cell is usually treated as a mass point located at
its center. When the planet travels very close to the center of the
cell, we get a gravitational singularity and the planet would suffer
extremely large gravity force. However, since the density is uniform
within a cell, the net force exerted on the planet should vanish because
of the symmetry of the cell. A softening length is always needed to
avoid the singularity, $\Phi_p = -G M_p/(|\vec{r}-\vec{r}_p|+\varepsilon)
$.

The softening length $\varepsilon$ could only reduce the amplitude of
the gravity impulses, however, it could not result in the real gravity
exerted on the planet. The choice of softening length is very tricky: a
small $\varepsilon$ could not reduce the singularity effectively, while
a large one would eliminate too much physical effect. It is usually set
to be a large fraction (e.g. $0.6-0.8$) of the scale height of the disk
or the Hill radius of the planet. However, in a low resolution grid, the
Hill radius only covers a few cells. Many local physical interactions
between the planet and disk would be concealed if we chose $\varepsilon$
compared to the Hill radius. To more reliably model the gravity felt by
the planet, we treat a single cell as a uniform area and the gravity
exerted on planet is an integration over this area, e.g. the gravity
force at the $\theta$ direction reads:
\begin{equation}
F_{\theta,i,j}=GM_p\sigma\int^{r_{i+\frac{1}{2}}}_{r_{i-\frac{1}{2}}}\int^{\theta_{j+\frac{1}{2}}}_{\theta_{j-\frac{1}{2}}}
\frac{r^2\sin(\theta-\theta_p)}{(r^2_p+r^2-2rr_p\cos(\theta-\theta_p)+\varepsilon)^{3/2}}drd\theta.
\end{equation}
$\varepsilon$ is now an integration softening parameter which is very
small. In our simulations, we set $\varepsilon=10^{-4}$ in dimensionless
units (the radius of the Roche lobe now is $\sim0.07$ and grid size is
$\sim0.01$). This treatment is performed in $5\times5$ cells around the
cell where the planet is located. The cells outside this $5\times5$
region are treated as point masses as usual.

\begin{figure}
\includegraphics[width=140mm]{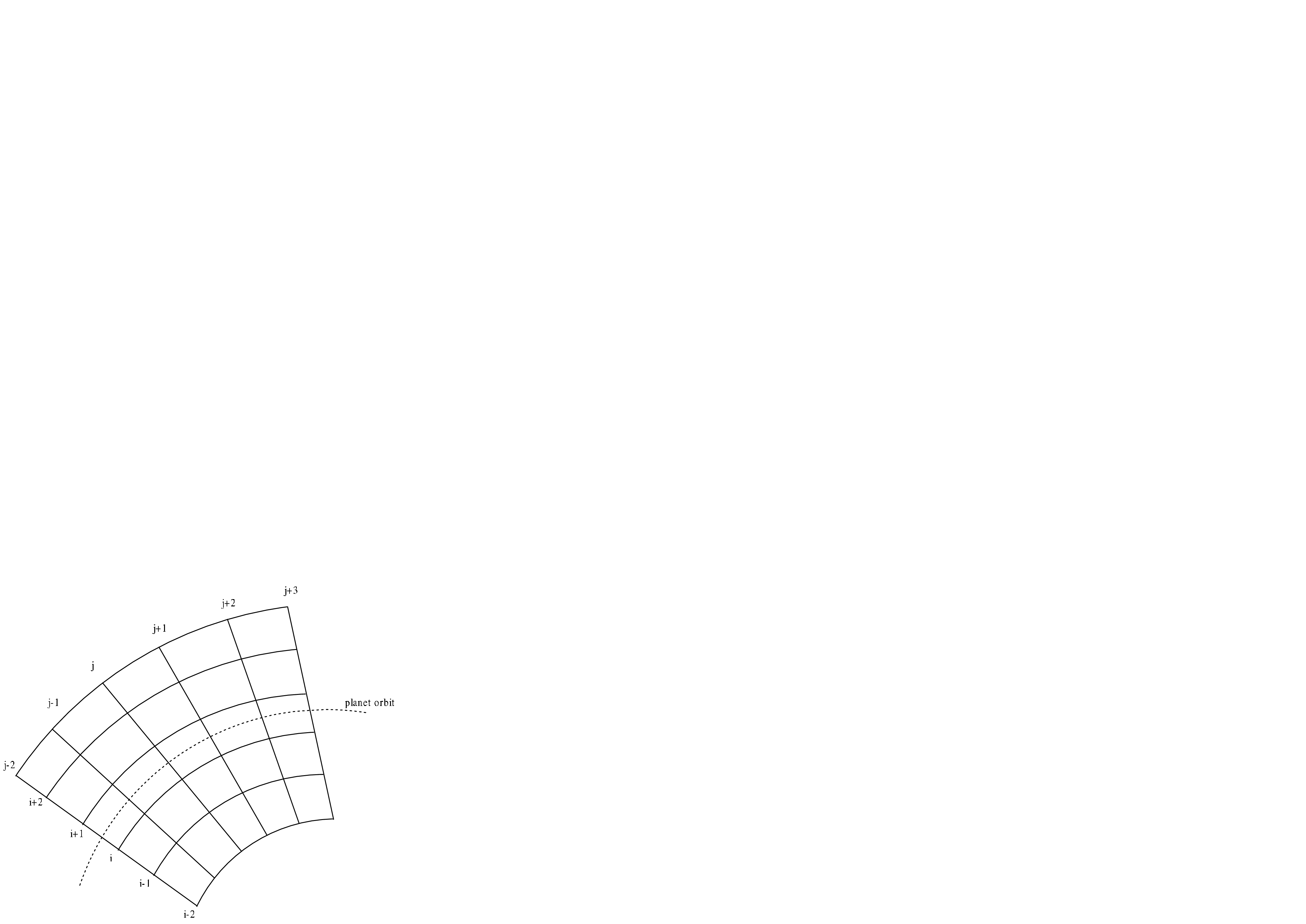}
\caption{A $5\times5$-grid in the vicinity of the planet. When we
calculate the torque exerted on the planet by this area, each cell is
treated as an uniform area instead of mass point. The dotted line shows
the track of the planet in the comparison tests.}\label{5x5cells}
\end{figure}

\begin{figure}
\includegraphics[width=140mm]{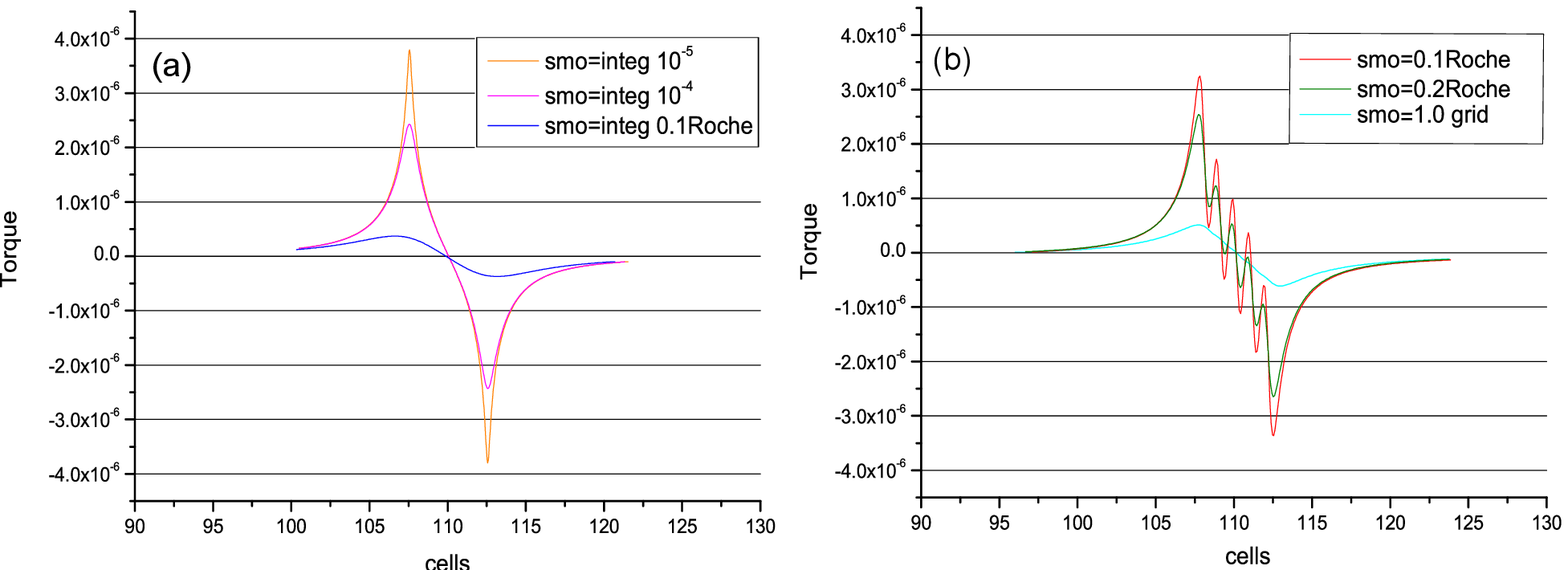}
\caption{A test of the gravity torques exerted on the planet under different treatments. The planet travels in a circular orbit and the disk is divided in to $256x512$ cells. The center of test area---a $5x5$-grid(\ref{5x5cells})---locates at $x=110$. The x-axis denotes the cell numbers(grid).
Panel (a): cells around the planet are treated as uniform areas. We perform integration over each of these cells to find the torques they exerted on the planet. The smoothing length used in the integration is shown in legend. $smo=integ 10^{-5}$ means the softening length used in the integration is $10^{-5}$ in our unit.
Panel (b): cells are treated as point mass. We assume the mass of a cell concentrates at its center. The gravity between the planet and a cell center is calculated with a softening length to avoid singularity. $smo=0.1 Roche$ means the softening length is one tenth of the initial Roche radius of the planet which is $\sim0.069$ in our units. $smo=1 gridsize$ means the softening length equals to the grid size. Treating cells as point masses usually results in large gravity impulses or over-smoothed gravity, while treating cells as area gives more smooth results and avoids any un-physical oscillations.}
\label{celltorque}
\end{figure}

A comparison of different treatments of gravity is performed. We set a
region of $5\times5$-grid whose surface density is uniform and
$\Sigma_0=1MMSN$. Outside this region the surface density is set to be
$0$ (fig. \ref{5x5cells}). As the planet travels through this region,
the gravity exerted on it should change smoothly and symmetrically from
the positive to the negative extrema, and vanishes at the center of this
area. The results are shown in fig. \ref{celltorque}.  It is clear
that the gravity is over-smoothed by the large $\varepsilon \sim 1 gridsize$ while the smaller one $\varepsilon=0.1-0.2 R_{Roche}$ introduces
nonphysical gravity impulses (panel (b) of Fig. \ref{celltorque}). Only the integration results with small $\varepsilon \sim 10^{-4\sim-5}$ could void the nonphysical gravity impulses (panel (a) of fig. \ref{celltorque}).

We also test the net torque of the whole disk under different treatments. The result is shown in fig. \ref{torque}. When we treat a cell of the disk as a point mass, the mutual gravity between the planet and the cell is very sensitive to the distance between them. When the planet travels through a high density cell and is very close to the cell center, its net torque will be dominated by this single cell. As the planet keeps passing by these point masses, the net torque exerted on it oscillates violently(the blue line in fig. \ref{torque}). On contrast, when we treat a cell as a continuous uniform area, the net gravity from the cell vanishes when the planet locates at the center. As a result, the net torque becomes more smooth and reliable(the red line in fig. \ref{torque}).

\begin{figure}
\includegraphics[width=140mm]{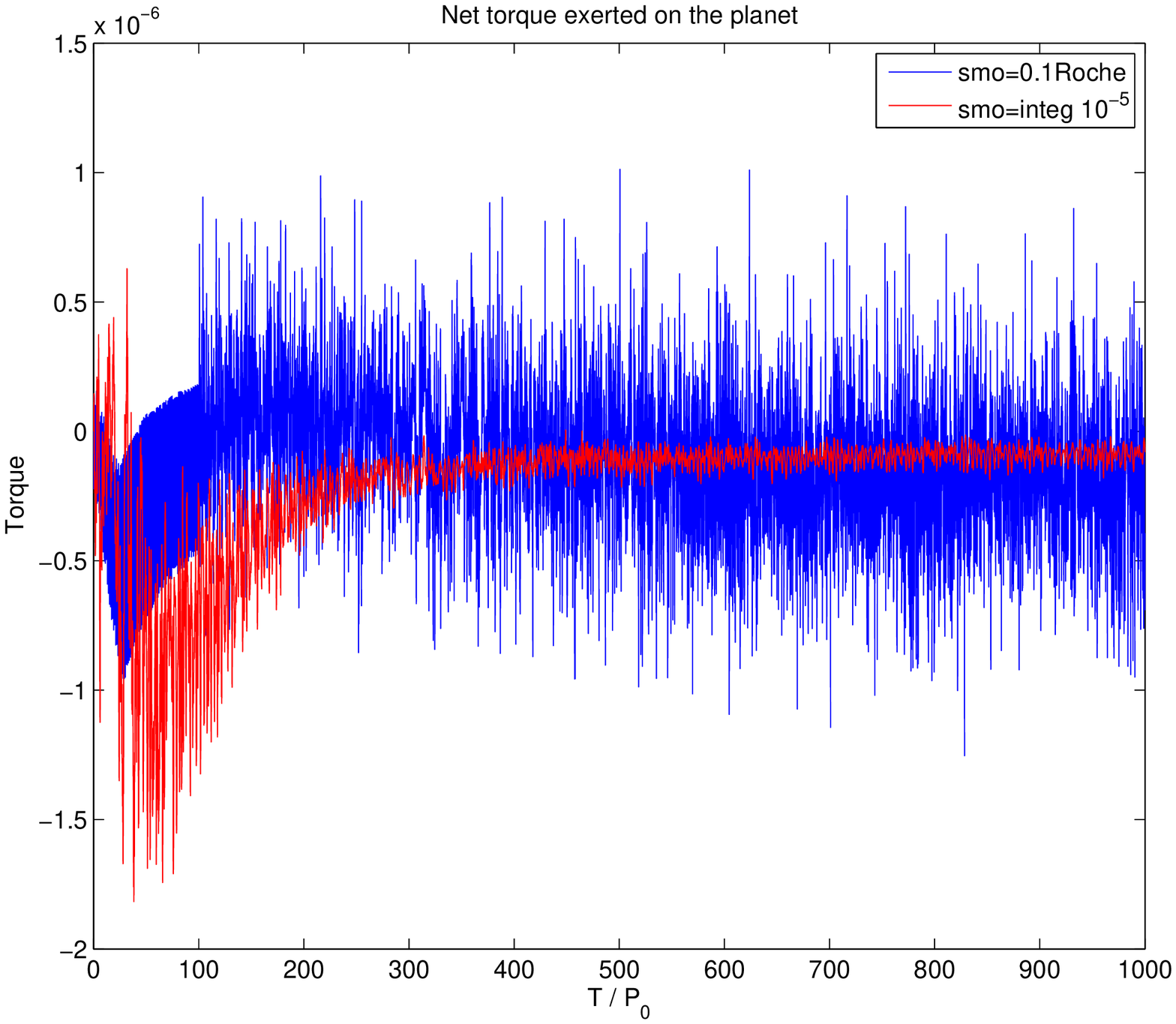}
\caption{Another test of the gravity torques exerted on the planet under different treatments. We test the net torques exerted on the planet by the whole disk. Blue solid line shows the torque when we treat the cells as point masses, while the red solid line shows the torque when the cells around the planet are treated as uniform areas. The large oscillations in the former torque are the results when the planet travels through some dense cells (close to a large point mass). It is clearly that the later treatment gives more reliable results.}
\label{torque}
\end{figure}

\section{Self-gravity force of the disk}\label{sg}

The self-gravitating effect of gas is included in the evolution of the
disk. As the density distribution is changing with time, the gravity
potential of the disk evolves and needs to be determined by solving the
Poisson equation at each time step: $\nabla^2\Phi_D = 4\pi G\Sigma$.
Integrating it over the disk in polar coordinates we get:
\begin{equation}
\Phi_D(r,\theta)=2G\int\int\frac{\Sigma(r',\theta')}{(r^2+r'^2-2rr'
\cos(\theta'-\theta))^{1/2}}r'dr'd\theta',
\end{equation}
However, solving this equation directly is very ``expensive'' even in
coarse resolution and the Fast Fourier Transform (FFT) method is one of
the best choices.

The self-gravity force exerting on each cell in the radial direction reads:
\begin{equation}
S_r(r,\theta)=-2G\int\int\frac{\Sigma(r',\theta')r'(r-r'\cos(\theta'-\theta))}
{(r^2+r'^2-2rr'\cos(\theta'-\theta))^{3/2}}dr'd\theta'.
\end{equation}
Note that the right hand of the above equation is the convolution of
$\Sigma(\vec{r}')$ and $K(\vec{r}-\vec{r}')$, where
\begin{equation}
K\equiv-2G\frac{(r-r'\cos(\theta'-\theta))}{(r^2+r'^2-2rr'\cos(\theta'-\theta))^{3/2}}.
\end{equation}
According to the 'convolution theorem' we can get $S_r$ by two Fourier
transforms($F$) and one reversed Fourier transform($F^{-1}$)(\cite{nr92}):
\begin{equation}
S_r=F^{-1}(F(\Sigma)F(K)).
\end{equation}
The kernel $K$ in fact does not change with time and only needs to be
calculated once at the beginning of the simulation. The self-gravity
force in the azimuthal direction can be obtained similarly. The detailed
introduction of this method can be found in many computational method
handbooks, e.g. ``Numerical recipes'' (\cite{nr92}).

To avoid the self-gravity potential being abruptly cut off at each
boundary of the disk, we add two buffer rings immediately outside the
boundaries. The width of each buffer ring is $0.3$ in our units and
their surface densities do not evolve with time. We integrate the radial
gravities of these two buffer rings and add them to the total gravity of
the disk.

\clearpage

\end{document}